\documentclass[letterpaper, 10 pt, conference]{ieeeconf}  

\IEEEoverridecommandlockouts                              

\usepackage{graphicx} 
\usepackage{amsmath} 
\usepackage{amssymb}  
\usepackage{booktabs}
\usepackage{wrapfig}
\usepackage{hyperref}
\usepackage{algorithm}
\usepackage{algpseudocode}
\usepackage{stfloats}

\title{\LARGE \bf
Neural Identification for Control
}

\author{Priyabrata Saha, Magnus Egerstedt, and Saibal Mukhopadhyay
\thanks{Authors are with School of Electrical and Computer Engineering,
Georgia Tech, Atlanta, USA.}
\thanks{Correspondence:  {\tt\small priyabratasaha@gatech.edu}}
\thanks{Code and models are available at 
\href{https://github.com/sahapriyabrata/NI4C}{https://github.com/sahapriyabrata/NI4C}}
}

\begin{document}

\maketitle
\thispagestyle{empty}
\pagestyle{empty}

\begin{abstract}
We present a new method for learning control law that stabilizes an unknown nonlinear dynamical system at an equilibrium point. We formulate a system identification task in a self-supervised learning setting that jointly learns a controller and corresponding stable closed-loop dynamics hypothesis. The input-output behavior of the unknown dynamical system under random control inputs is used as the supervising signal to train the neural network-based system model and the controller. The proposed method relies on the Lyapunov stability theory to generate a stable closed-loop dynamics hypothesis and corresponding control law.  We demonstrate our method on various nonlinear control problems such as n-link pendulum balancing and trajectory tracking, pendulum on cart balancing, and wheeled vehicle path following.
\end{abstract}

\section{Introduction}
\label{sec:intro}

Designing a controller to stabilize a nonlinear dynamical system has been an active area of research for decades. Classical approaches 
generally involve linearization of the system. In \textit{Jacobian linearization}, the system dynamics is linearized within a small neighborhood around the equilibrium point and then a linear control, e.g., \textit{linear-quadratic regulators} (LQR \cite{kwakernaak1972linear}), is applied to stabilize the system. This method is not suitable when a large region of operation is required or the system involves `hard nonlinearities' that do not allow linear approximation \cite{slotine1991applied}. \textit{Feedback linearization}, on the other hand, constructs a nonlinear controller by canceling system nonlinearities with algebraic transformations so that the closed-loop system takes a (fully or partially) linear form \cite{slotine1991applied}. Another powerful method for nonlinear controller design is the method of \textit{control-Lyapunov function} \cite{haddad2011nonlinear,sontag1989universal}.
However, the problem of constructing a Lyapunov function is very hard in general \cite{giesl2015review}.

Neural networks have been explored for designing control of nonlinear systems. However, most of the prior works are focused on control-affine systems \cite{lewis1996neural, gao2006neural, dinh2014dynamic}. Neural network-based control design for nonaffine systems generally assumes the system to be in Brunovsky form \cite{ge2003neural} or pure-feedback form \cite{ge2002adaptive}. Recently, Patan and Patan \cite{patan2020neural} used a neural network-based iterative learning control for an unknown nonlinear plant. They model the system dynamics using a neural network and develop a neural controller using the learned model of the system dynamics.

The aforementioned works use neural networks with a single hidden layer and saturating activation functions.
However, it has been observed that a deep network requires less number of parameters in comparison with its shallow counterpart to approximate a composite function with similar accuracy \cite{mhaskar2017and}. Hence, there has been a growing interest in utilizing deep neural networks (DNNs) to design controllers for nonlinear systems. Chang et al. \cite{chang2019neural} used DNN to learn a Lyapunov function and design a controller for a complex system when the dynamics is known \textit{a-priori}. Authors trained a neural Lyapunov function and update the parameters of an initial LQR controller by minimizing a Lyapunov-constrained loss function. Taylor et al. \cite{taylor2019episodic, taylor2019control} used DNN to learn the uncertain part of a partially known system and iteratively update the corresponding Lyapunov function to improve an existing controller. Bhatt et al. \cite{bhatt2019context} used a neural network to predict trajectory execution error of industrial manipulator and used that to design a context-dependent compensation scheme.

Most of the deep learning methods mentioned above require some knowledge of the system dynamics. However, in many real problems, the system dynamics is unknown. The need for \textit{automatic synthesis of control algorithms} for unknown systems has long been recognized in formal control \cite{hjalmarsson2005experiment,geversaa2005identification}. We present a novel deep learning approach, referred to as the \textit{neural identification for control}, to design controller for an unknown nonlinear dynamical system that couples the expressive power of neural networks with concepts from classical feedback linearization and Lyapunov stability theory. We consider the following constraints: (1) only observations of inputs and states, rather than any (fully or partially) known model, of the system is available, (2) a stable closed-loop response of the system is not available for supervision and no pre-existing controller is available for initialization.

Our approach is motivated by the extensive research in formal control on \textit{identification for control} \cite{hjalmarsson2005experiment,geversaa2005identification}. 
We recognize that directly learning an accurate model of an unknown nonlinear dynamics can be arbitrarily complex. However, in identification for control, the closed-loop performance of the learned controller is of primary concern, rather than an accurate model of system dynamics \cite{geversaa2005identification}. We leverage this observation to formulate a system identification task in a self-supervised setting that learns the control law to stabilize a system at an equilibrium point without explicitly seeking an accurate model of the unknown dynamics.
We analytically show that a neural network can be utilized to generate a stable closed-loop dynamics hypothesis exploiting Lyapunov stability theory. We use that hypothesis to learn the control law from the \textit{observations of the input-output behavior} of the system. 
We illustrate the proposed method on several nonlinear control problems, namely, balancing an n-link pendulum and trajectory tracking, balancing a pendulum on a cart, and tracking the path of a wheeled vehicle. 

Training neural networks for unknown systems requires a large amount of data which can be challenging for safety-critical systems, particularly, when training data must be collected from a broad region of operation. We propose to address this problem using an iterative approach, where we initially collect training data from a known small safe region around the equilibrium point and learn an initial controller. Next, we use the learned controller to collect data over a wider region to update the controller. Using the example of an n-link pendulum, we show that our method can learn a controller that yields a \textit{region of attraction} (ROA) beyond the training domain making such an iterative approach possible.
\section{Problem Statement and Preliminaries} 
\label{sec:prelim}
 
\subsection{Problem Statement}
Consider a time-invariant controlled dynamical system of the form
\begin{equation}
    \frac{d\mathbf{x}}{dt} = f (\mathbf{x}, \mathbf{u}), \quad \mathbf{x(0)} = \mathbf{x}_0,
    \label{eqn:controlled_system}
\end{equation}
where $\mathbf{x}(t) \in \mathcal{X} \subset \mathbb{R}^n$ and $\mathbf{u}(t) \in \mathcal{U} \subset \mathbb{R}^m$ are the system state and control input, respectively, at time $t$. $f: \mathcal{X} \times \mathcal{U} \rightarrow  \mathbb{R}^n$ is some unknown nonlinear function. When the control input is zero, i.e. $\mathbf{u} = \mathbf{0}$, we call the system, $\frac{d\mathbf{x}}{dt} = f (\mathbf{x}, \mathbf{0}) = f_0 (\mathbf{x})$, an autonomous dynamical system. For the dynamical system of (\ref{eqn:controlled_system}), assuming it is stabilizable, we consider the following problem. 

\textbf{Problem.} \textit{Learn a feedback control law $\mathbf{u} = \pi(\mathbf{x})$ for an unknown time-invariant dynamical system of (\ref{eqn:controlled_system})
that makes the corresponding closed-loop 
dynamics $\frac{dx}{dt} = f(\mathbf{x}, \pi(\mathbf{x}))$ asymptotically stable at an equilibrium point $\mathbf{x}_e$, i.e., $\forall \ \mathbf{x}(0) \in \mathcal{X}_{\pi}, \lim_{t\to\infty} \| \mathbf{x}(t)\| = \mathbf{x}_e$, where $\mathcal{X}_{\pi}$ is the ROA of the closed-loop system under the control law $\pi$.} 

\subsection{Lyapunov stability}
Stability of a dynamical system at equilibrium points is usually characterized using the method of Lyapunov. Suppose the origin $\mathbf{x} = \mathbf{0}$ be an equilibrium point for a dynamical system $\frac{d\mathbf{x}}{dt} = h (\mathbf{x})$, where $h : \mathcal{X} \rightarrow  \mathbb{R}^n$ is a locally Lipschitz map. Let $V : \mathcal{X} \rightarrow \mathbb{R}$ be a continuously differentiable function such that
\begin{equation}
    V(\mathbf{0}) = 0, \quad \text{and} \quad  V(\mathbf{x}) > 0 \quad \forall \ \mathbf{x} \in \mathcal{X} \setminus \{\mathbf{0}\},
\end{equation}
and the time derivative of $V$ along the trajectories
\begin{equation}
    \frac{dV}{dt} = \nabla V(\mathbf{x})^T \frac{d\mathbf{x}}{dt} = \nabla V(\mathbf{x})^T h(\mathbf{x}) \leq 0 \quad \forall \ \mathbf{x} \in \mathcal{X}. 
\end{equation}
Then, the origin is stable and $V$ is called a Lyapunov function. Moreover, if there exist constants $k_1 > 0, k_2 > 0,$ and $\alpha > 0$ such that
\begin{align}
    k_1 \| \mathbf{x}\|^2 &\leq V(\mathbf{x}) \leq k_2 \| \mathbf{x}\|^2, \nonumber \\
    \text{and} \quad \nabla V(\mathbf{x})^T h(\mathbf{x}) &\leq - \alpha \| \mathbf{x}\|^2 \quad \forall \ \mathbf{x} \in \mathcal{X},
\end{align}
then the origin is exponentially stable 
\cite{khalil2002nonlinear}.
\section{Proposed Learning Approach}
\label{sec:learning}

\begin{figure*}
  \centering
  \includegraphics[width=0.85\linewidth]{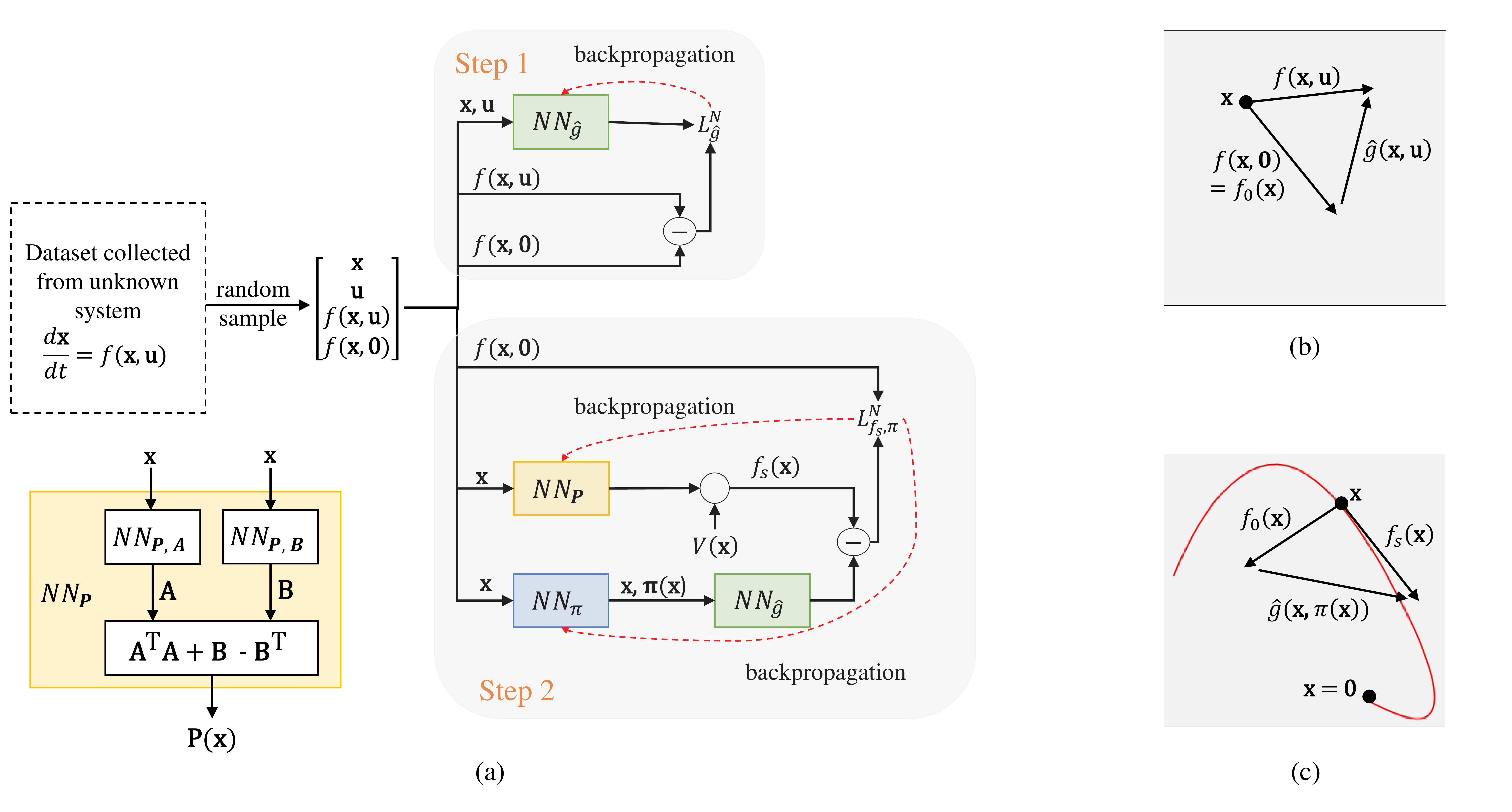}
  \vspace{-5pt}
  \caption{The two steps of the learning process. (a, b) First, the effect of control input on the dynamics is learned by the neural network $\text{NN}_{\hat g}$ using observed data of autonomous dynamics $f_0$ and dynamics $f$ under random control input. (a, c) Next, the stable closed-loop dynamics hypothesis $f_s$ and corresponding control law $\pi$ are learned jointly by neural networks $\text{NN}_{\mathbf{P}}$ and $\text{NN}_{\pi}$, respectively, using observed data of autonomous dynamics $f_0$. Structure of $\text{NN}_{\mathbf{P}}$ is shown separately in (a). The red curve in (c) shows a stable trajectory hypothesis converging to the origin and $f_s$ is its current component at state $\mathbf{x}$.}
  \label{fig:learning}
  \vspace{-5pt}
\end{figure*}

We formulate a system identification task to learn a control law to stabilize a nonlinear system without explicitly seeking an accurate model of the system dynamics. Our learning method involves two steps which are designed based on the following rearrangement of the controlled dynamics of (\ref{eqn:controlled_system}):
\begin{align}
    \frac{d\mathbf{x}}{dt} = f (\mathbf{x}, \mathbf{u}) &= f(\mathbf{x}, \mathbf{0}) + (f(\mathbf{x}, \mathbf{u}) - f(\mathbf{x}, \mathbf{0})) \nonumber \\
    & = f_0(\mathbf{x}) + g(\mathbf{x}, \mathbf{u}) , \quad \mathbf{x(0)} = \mathbf{x}_0
    \label{eqn:considered_system}
\end{align}
Based on (\ref{eqn:considered_system}), the two steps of learning methods are defined as follows.
\begin{itemize}
	\item First, we train a neural network to learn the effect of control input (i.e. an approximate model for $g(\mathbf{x}, \mathbf{u})$).
	
	\item Next, we train two neural networks jointly. One network learns to generate a stable closed-loop dynamics hypothesis $f_s(\mathbf{x})$ and the other one learns a control law $\pi(\mathbf{x})$ that drives the system toward $f_s$. 
\end{itemize}
Figure \ref{fig:learning}(a) shows these two steps schematically. Note, the neural network approximation does not require the function $g(\mathbf{x}, \mathbf{u})$ to be affine with respect to control input $\mathbf{u}$; therefore, our method can be used for non-affine systems in general. 

We assume a self-supervised learning setting where all the training data, i.e. the observations of input $\mathbf{u}$, state $\mathbf{x}$ and dynamics $f(\mathbf{x}, \mathbf{u})$, are collected beforehand within the region of operation $\mathcal{X}$ and $\mathcal{U}$.

\subsection{Learning the effect of control input}
To learn the effect of control input in dynamics i.e., a neural network approximation of function $g$ of (\ref{eqn:considered_system}), we use the difference between autonomous dynamics and dynamics subjected to a random nonzero control input (Figure \ref{fig:learning}(b)). We sample state $\mathbf{x}$ and control input $\mathbf{u}$ from $\mathcal{X} \times \mathcal{U}$ with joint distribution $p_{\mathbf{xu}}$ and train a neural network for $\hat g$ (denotes an approximation of $g$) to minimize the loss 
\begin{equation}
    L_{\hat g} =  \mathbb{E}_{\mathbf{x},\mathbf{u} \ \sim \ p_{\mathbf{xu}}(\mathcal{X} \times \mathcal{U})} \quad \| f (\mathbf{x}, \mathbf{u}) - f_0(\mathbf{x}) - \hat g (\mathbf{x}, \mathbf{u})\|^2
\end{equation}
It is important to note that the values of $f$ and $f_0$ are obtained from observations of the system (training data), not computed from a known model. Specifically in experiment, $N$ samples $(\mathbf{x}_i, \mathbf{u}_i), i \in \{1, 2, \cdots, N\}$ are drawn from $\mathcal{X} \times \mathcal{U}$ according to $p_{\mathbf{xu}}$ and the model is trained to minimize the following empirical loss. 
\begin{equation}
    L_{\hat g}^N = \frac{1}{N} \sum_{i=1}^N \| f (\mathbf{x}_i, \mathbf{u}_i) - f(\mathbf{x}_i, \mathbf{0}) - \hat g (\mathbf{x}_i, \mathbf{u}_i)\|^2
    \label{eqn:loss1}
\end{equation}

\subsection{Learning the stable dynamics hypothesis and control law}
We propose the following: 
\vspace{2mm}

\textit{A control law $\mathbf{u} = \pi(\mathbf{x})$ that stabilizes the system at the origin and the corresponding stable closed-loop dynamics $f_s$ can be learned jointly by minimizing the loss}
\begin{equation}
    L_{f_s, \pi} = \mathbb{E}_{\mathbf{x} \ \sim \ p_{\mathbf{x}}(\mathcal{X})} \quad \| f_0 (\mathbf{x}) + \hat g (\mathbf{x}, \pi(\mathbf{x})) - f_s(\mathbf{x}) \|^2 \ ,
    \label{eqn:loss2_actual}
\end{equation}
\textit{where the $\mathbf{x}$ is a random variable over the state space $\mathcal{X}$ with a distribution $p_\mathbf{x}$.}

Effectively, we propose to train two neural networks jointly: one for hypothesizing a stable closed-loop dynamics $f_s$ directly from system state (since closed-loop dynamics can be described using only the system state) and, the other neural network for generating a control input $\pi(\mathbf{x})$. The estimated effect of the control input $\hat g(\mathbf{x}, \pi(\mathbf{x}))$ when added to the autonomous dynamics $f_0(\mathbf{x})$ should match with the hypothesis $f_s$ (Figure \ref{fig:learning}(c)). The difference between the hypothesized closed-loop behavior $f_s$ and the estimated behavior of the actual system, subjected to the policy $\pi$, is minimized using the loss function (\ref{eqn:loss2_actual}).

The true value of autonomous dynamics $f_0$ is obtained from the system by applying zero control input. Specifically in experiment, $N$ samples $\mathbf{x}_i, i \in \{1, 2, \cdots, N\}$ are drawn from $\mathcal{X}$ according to $p_{\mathbf{x}}$ and the neural networks for $f_s$  and $\pi$ are trained to minimize the following empirical loss. 
\begin{equation}
    L_{f_s, \pi}^N = \frac{1}{N} \sum_{i=1}^N \| f (\mathbf{x}_i, \mathbf{0}) + \hat g (\mathbf{x}_i, \pi(\mathbf{x}_i)) - f_s(\mathbf{x}_i) \|^2
    \label{eqn:loss2}
\end{equation}

A key component of our approach is a neural network that generates a stable dynamics hypothesis. This network is designed in such a way that the hypothesized closed-loop dynamics is proportional to the negative gradient of a Lyapunov energy function. The proportionality matrix, say $\mathbf{P} \in \mathbb{R}^{n \times n}$, should be such that $\mathbf{v}^T \mathbf{P}(\mathbf{x}) \mathbf{v} > 0$, for any $\mathbf{v} \in \mathbb{R}^n$. We first form a matrix $\mathbf{P}$ ensuring $\mathbf{v}^T \mathbf{P}(\mathbf{x}) \mathbf{v} \geq 0$, for any $\mathbf{v} \in \mathbb{R}^n$, and show how it can be used to generate a stable dynamics. However, exponential stability requires strict inequality ($\mathbf{v}^T \mathbf{P}(\mathbf{x}) \mathbf{v} > 0$). For that purpose, we use the technique proposed in \cite{kolter2019learning} that ensures a positive decay in Lyapunov energy resulting in exponential stability. The overall method of designing a neural network to generate a stable dynamics hypothesis is developed using the following result. \\ 
\textbf{Theorem 1.} \textit{Let $V : \mathcal{X} \rightarrow \mathbb{R}$ be a function defined by}
\begin{equation}
    V(\mathbf{x}) = \mathbf{x}^T \mathbf{Q} \mathbf{x}, \quad \mathbf{x} \in \mathcal{X} \subset \mathbb{R}^n,
     \label{eqn:lyapunov}
\end{equation}
\textit{where $\mathbf{Q} \in \mathbb{R}^{n \times n}$ is a positive definite matrix. Suppose $\mathbf{P} : \mathcal{X} \rightarrow \mathbb{R}^{n \times n}$ be a function such that}
\begin{equation}
    \mathbf{P}(\mathbf{x}) = \mathbf{A}(\mathbf{x})^T\mathbf{A}(\mathbf{x}) + \mathbf{B}(\mathbf{x}) - \mathbf{B}(\mathbf{x})^T, 
    \label{eqn:Px}
\end{equation}
\textit{where $\mathbf{A}(\mathbf{x}) \in \mathbb{R}^{l \times n}$ and $\mathbf{B}(\mathbf{x}) \in \mathbb{R}^{n \times n}$ are some arbitrary functions of $\mathbf{x}$.}
\textit{Then, the dynamics defined by}
\begin{equation}
    \frac{d\mathbf{x}}{dt} = -\mathbf{P}(\mathbf{x}) \nabla V(\mathbf{x}), \quad \mathbf{x(0)} = \mathbf{x}_0,
    \label{eqn:stable}
\end{equation}
\textit{is stable at the origin $\mathbf{x}=\mathbf{0}$. Moreover, for any constant $\alpha > 0$, the dynamics}
\begin{align}
\frac{d\mathbf{x}}{dt} &= f_s (\mathbf{x}) \nonumber \\ &= - \mathbf{P}(\mathbf{x})\nabla V(\mathbf{x}) - \frac{\text{ReLU}\big(- W(\mathbf{x}) + \alpha V(\mathbf{x})  \big)}{\|\nabla V(\mathbf{x}) \|^2} \nabla V(\mathbf{x}), \nonumber \\ & \hspace{60mm} \mathbf{x(0)} = \mathbf{x}_0,
   \label{eqn:exp_stable}
\end{align}
\textit{where} $W(\mathbf{x}) = \nabla V(\mathbf{x})^T \mathbf{P}(\mathbf{x}) \nabla V(\mathbf{x}),$ \textit{and} $\text{ReLU}(z) = \text{max}(\\0, z), z \in \mathbb{R},$
\textit{is exponentially stable at the origin $\mathbf{x}=\mathbf{0}$.} \\ \\
\textit{Proof.} It can be proved straightforwardly by applying \textit{Lyapunov's stability theorem} (section \ref{sec:prelim}.B) and  
the fact that for any $\mathbf{v} \in \mathbb{R}^n$, $\mathbf{v}^T \mathbf{P}(\mathbf{x}) \mathbf{v} \geq 0$ where $\mathbf{P}(\mathbf{x})$ is given by (\ref{eqn:Px}).

By definition $V$ in (\ref{eqn:lyapunov}) is continuously differentiable and we have
\begin{equation}
    V(\mathbf{0}) = 0, \quad \text{and} \quad  V(\mathbf{x}) > 0 \quad \forall \ \mathbf{x} \in \mathcal{X} \setminus \{\mathbf{0}\}
\end{equation}
Now, for the dynamics of (\ref{eqn:stable})  we get
\begin{align}
    \frac{dV}{dt} &= \nabla V(\mathbf{x})^T \frac{d\mathbf{x}}{dt} \nonumber \\ &= - \nabla V(\mathbf{x})^T \mathbf{P}(\mathbf{x}) \nabla V(\mathbf{x}) \leq 0 \quad \forall \ \mathbf{x} \in \mathcal{X}. 
\end{align}
Therefore, according to Lyapunov's stability theorem, (\ref{eqn:stable}) is stable at the origin $\mathbf{x} = \mathbf{0}$.

From the definition of $V$ in (\ref{eqn:lyapunov}), we have
\begin{equation}
   \lambda_{min}(\mathbf{Q}) \|\mathbf{x}\|^2 \leq V(\mathbf{x}) \leq \lambda_{max}(\mathbf{Q}) \|\mathbf{x}\|^2,
\end{equation}
where, $\lambda_{min}(\mathbf{Q})$ and $\lambda_{max}(\mathbf{Q})$ denote the smallest and largest eigenvalues, respectively, of $\mathbf{Q}$ and have positive values since the matrix $\mathbf{Q}$ is positive definite.
Now, $W(\mathbf{x}) = \nabla V(\mathbf{x})^T \mathbf{P}(\mathbf{x}) \nabla V(\mathbf{x}) \geq 0$ and for the dynamics of (\ref{eqn:exp_stable}), we have
\begin{align}
    \frac{dV}{dt} &= \nabla V(\mathbf{x})^T \frac{d\mathbf{x}}{dt} \nonumber\\
                  &= \begin{cases}
                          - \nabla V(\mathbf{x})^T \mathbf{P}(\mathbf{x}) \nabla V(\mathbf{x}), \\ \hspace{37mm} if \quad W(\mathbf{x}) \geq \alpha V(\mathbf{x}) \\
                          - \nabla V(\mathbf{x})^T \mathbf{P}(\mathbf{x}) \nabla V(\mathbf{x}) - (-W(\mathbf{x}) + \alpha V(\mathbf{x})), \\ \hspace{52mm} otherwise
                     \end{cases} \nonumber \\
                 &= \begin{cases}
                       -W(\mathbf{x}), & if \quad W(\mathbf{x}) \geq \alpha V(\mathbf{x}) \\
                       - \alpha V(\mathbf{x}), & otherwise
                    \end{cases}
    \label{eqn:decay}                    
\end{align}
Equation (\ref{eqn:decay}) implies
\begin{equation}
    \frac{dV}{dt} \leq - \alpha V(\mathbf{x}) \leq - \alpha \lambda_{min}(\mathbf{Q}) \|\mathbf{x}\|^2 < 0 \quad \forall \ \mathbf{x} \in \mathcal{X}
\end{equation}
Hence, according to Lyapunov's stability theorem, (\ref{eqn:exp_stable}) is exponentially stable at the origin $\mathbf{x} = \mathbf{0}$.    \hspace{2cm} $\blacksquare$

\subsection{Discussion}
\textbf{Choice of $\mathbf{P}(\mathbf{x})$.} It is important to note that the matrix  $\mathbf{A}(\mathbf{x})^T\mathbf{A}(\mathbf{x})$ in (\ref{eqn:Px}) is positive semi-definite, i.e,  $\mathbf{v}^T \mathbf{A}(\mathbf{x})^T\mathbf{A}(\mathbf{x}) \mathbf{v} \geq 0$, for any $\mathbf{v} \in \mathbb{R}^n$ and therefore, if we use $\mathbf{P}(\mathbf{x}) = \mathbf{A}(\mathbf{x})^T\mathbf{A}(\mathbf{x})$ in (\ref{eqn:stable}), the dynamics would be stable. However, $\mathbf{A}(\mathbf{x})^T\mathbf{A}(\mathbf{x})$ is a symmetric matrix which is not a necessary constraint for the stability matrix of a generic stable system. Therefore, we relax that constraint by adding a skew-symmetric matrix $\mathbf{B}(\mathbf{x}) - \mathbf{B}(\mathbf{x})^T$ in (\ref{eqn:Px}). 

\textbf{Exponential stability.} The equation (\ref{eqn:stable}) represents a dynamics that is stable at the origin. We follow the methods developed in \cite{kolter2019learning} to generate the exponentially stable dynamics hypothesis $f_s(\mathbf{x})$, shown in (\ref{eqn:exp_stable}), by subtracting a component in the direction of gradient of $V$. As shown in \cite{kolter2019learning}, this ensures a positive decay in Lyapunov energy
($V$) along the trajectory which in-turn guarantees exponential stability. 

\textbf{Differences with \cite{kolter2019learning}}. In \cite{kolter2019learning}, authors considered the problem of modeling a stable dynamical system using DNN. They use the trajectory data from a given stable system to train a model that replicates the stable behavior of the system. 
The model is trained using a loss that compares the dynamics obtained from the neural network with the true stable behavior of the underlying system.  
On the other hand, we consider the problem of learning a control law to stabilize an unstable system. Therefore, we do not have any true stable behavior to compare against in the loss function. Rather, we only use the unstable trajectory data to define a loss function for joint learning of the stable dynamics hypothesis and control law.
The problem of learning a Lyapunov function, a control law and a stable dynamics hypothesis jointly from unstable trajectories is very hard due to the local nature of stochastic gradient descent. Therefore, we use a specific type of Lyapunov function, namely quadratic Lyapunov function (defined by (\ref{eqn:lyapunov})), and constrain the stable dynamics hypothesis to be proportional to the gradient of the Lyapunov function. The positive definite matrix $\mathbf{Q}$ (in (\ref{eqn:lyapunov})) is a hyperparmeter and requires manual tuning. Choosing the values of $\mathbf{Q}$ is similar to selecting the cost matrices for LQR and depends on the prioritization of the state variables as per requirement. The value of the decay constant (of Lyapunov energy) $\alpha$ should be decided based on the requirement of underdamped or overdamped control; we show the effect of $\alpha$ in closed-loop system response in the experimental section. 

\begin{figure}[t]
  \centering
  \includegraphics[width=0.85\linewidth]{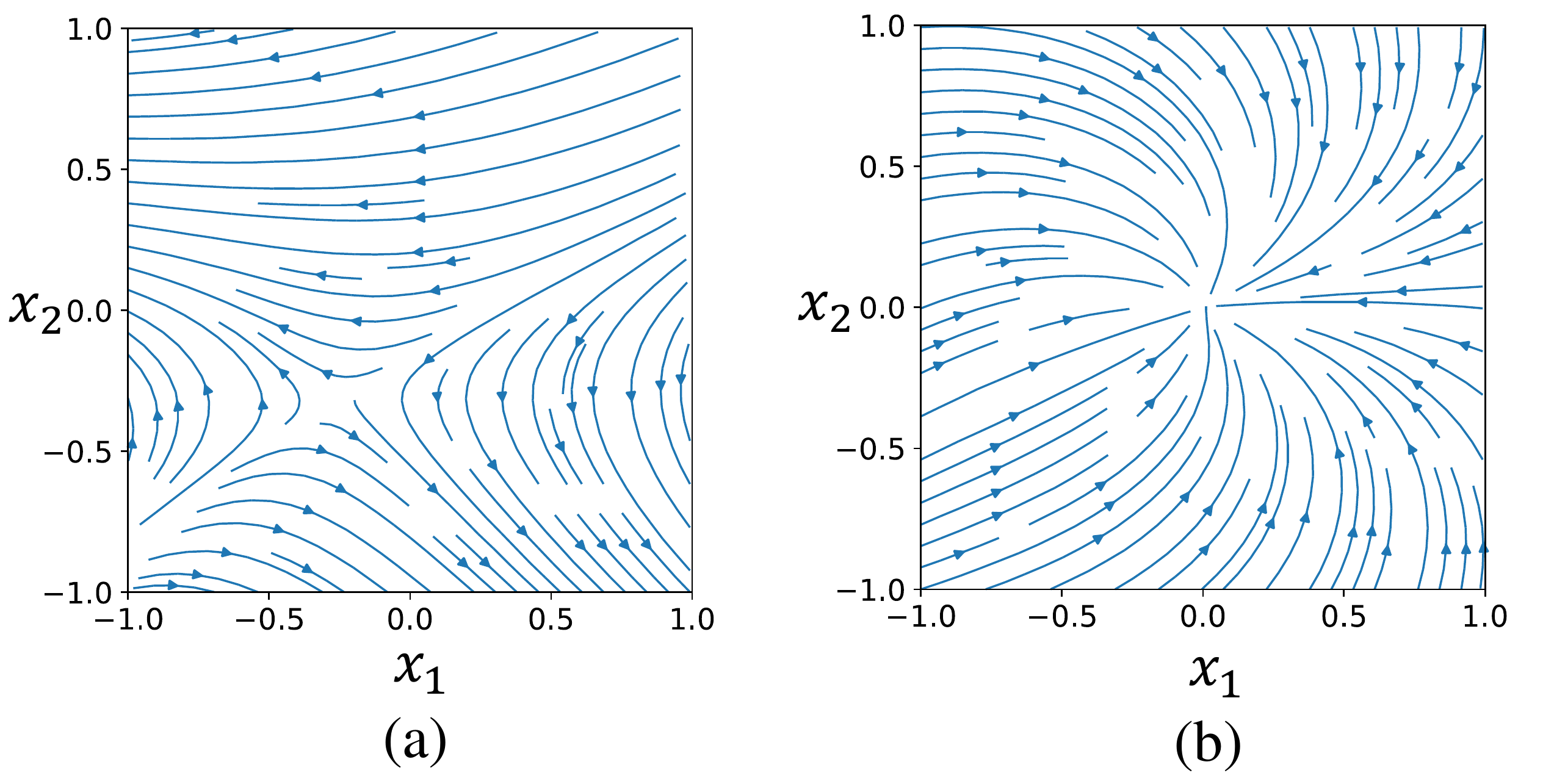}
  \vspace{-5pt}
  \caption{Phase portraits of a dynamics generated using neural network for a two-dimension system with state $\mathbf{x} = [x_1 , x_2]^T$. (a): Phase portrait of dynamics generated using a standard neural network with random weights. (b): Phase portrait of dynamics generated using the neural network (with random weights) designed according to equation (\ref{eqn:exp_stable}).}
  \label{fig:random}
 \vspace{-5pt}
\end{figure}

\subsection{Implementation} We use neural networks $\text{NN}_{\hat g}$, $\text{NN}_{\mathbf{P}}$, and $\text{NN}_{\pi}$ to represent the functions $\hat g$, $\mathbf{P}$ and $\pi$, respectively. The neural network $\text{NN}_{\mathbf{P}}$ is designed based on (\ref{eqn:Px}). The output neurons of a neural network constitute the entries of matrices $\mathbf{A}$ and $\mathbf{B}$, which are connected according to (\ref{eqn:Px}) to provide the final output $\mathbf{P}$ (Figure \ref{fig:learning}(a)). According to \textbf{Theorem 1}, for any arbitrary choice of matrices $\mathbf{A}$ and $\mathbf{B}$, we can get a stable dynamics using (\ref{eqn:exp_stable}). Therefore, the outputs of any neural network with random weights (even without any training) can be used as the elements of matrices $\mathbf{A}$ and $\mathbf{B}$ to generate a stable dynamics. Figure \ref{fig:random} shows that the neural network designed using (\ref{eqn:exp_stable}) inherently generates globally stable dynamics, i.e. all trajectories approaches origin, even without any training. However, by training this neural network with trajectory data of an unstable system, we enforce the generated stable dynamics hypothesis to be compatible with that underlying system.

First, we train $\text{NN}_{\hat g}$ using the loss function of (\ref{eqn:loss1}). Next, we train $\text{NN}_{\mathbf{P}}$, and $\text{NN}_{\pi}$ jointly using the loss function of (\ref{eqn:loss2}).
The overall procedure is summarized in \textbf{Algorithm 1}. During evaluation of the closed-loop performance, we only need the output of $\text{NN}_{\pi}$ to generate the control signal $\mathbf{u}$. 

\begin{algorithm}
  \caption{Neural Identification for Control}
  \begin{algorithmic}[1]
     \State  \textbf{input}: Black-box controlled dynamical system $f$, State space $\mathcal{X}$ and input space $\mathcal{U}$, decay constant $\alpha$, positive-definite matrix $\mathbf{Q} \in \mathbb{R}^{n \times n}$ 
     \State \ 
     \State Arbitrarily initialize the neural networks $\text{NN}_{\hat g}$, $\text{NN}_{\mathbf{P}}$, and $\text{NN}_{\pi}$
     \State \ 
     \Repeat \Comment Training $\text{NN}_{\hat g}$
     \State $\mathbf{x}, \mathbf{u} \sim p_{\mathbf{x}, \mathbf{u}}(\mathcal{X} \times \mathcal{U})$ \Comment \parbox[t]{.4\linewidth}{Sample batch of states and control inputs} 
     \State $\hat g(\mathbf{x}, \mathbf{u}) \leftarrow \text{NN}_{\hat g}(\mathbf{x}, \mathbf{u})$ \Comment \parbox[t]{.4\linewidth}{Forward pass the neural network}
     \State Get $f(\mathbf{x}, \mathbf{u})$ and $f(\mathbf{x}, \mathbf{0})$
     \State Compute the loss $L_{\hat g}^N$ using (\ref{eqn:loss1})
     \State \parbox[t]{0.9\linewidth}{Update the parameters of $\text{NN}_{\hat g}$ by backpropagating $L_{\hat g}^N$ and using SGD}
     \Until convergence
     \State \ 
     \Repeat \Comment Training $\text{NN}_{\mathbf{P}}$, $\text{NN}_{\pi}$
     \State $\mathbf{x} \sim p_{\mathbf{x}}(\mathcal{X})$ \Comment Sample batch of states
     \State $\mathbf{P}(\mathbf{x}), \  \pi(\mathbf{x}) \leftarrow \text{NN}_{\mathbf{P}}(\mathbf{x}), \  \text{NN}_{\pi}(\mathbf{x})$ \Comment \parbox[t]{.25\linewidth}{Forward pass the neural networks}
     \State $\nabla V(\mathbf{x}) \leftarrow 2\mathbf{Qx}$ \Comment \parbox[t]{.5\linewidth}{Compute the gradient of the Lyapunov function}
     \State Compute $f_s(\mathbf{x})$ according to (\ref{eqn:exp_stable})
     \State $\hat g(\mathbf{x}, \pi(\mathbf{x})) \leftarrow \text{NN}_{\hat g}(\mathbf{x}, \pi(\mathbf{x}))$
     \State Get $f(\mathbf{x}, \mathbf{0})$
     \State Compute the loss $L_{f_s, \pi}^N$ using (\ref{eqn:loss2})
     \State \parbox[t]{.9\linewidth}{Update the parameters of $\text{NN}_{\mathbf{P}}$ and $\text{NN}_{\pi}$ by backpropagating $L_{f_s, \pi}^N$ and using SGD}
     \Until convergence
  \end{algorithmic}
\end{algorithm}
\section{Simulation Results}
\label{sec:experiments}

We provide simulation results using the proposed method on several nonlinear control problems, namely, balancing an n-Link pendulum and trajectory tracking, balancing a pendulum on a cart, and tracking the path of a wheeled vehicle. We first describe the general simulation settings used for all the examples and then provide example specific details. Finally, we discuss the results. 

\subsection{Simulation settings}

\textbf{Verification of the learned controller.} We evaluate a learned control law $\pi$ by estimating the corresponding ROA $\mathcal{X}_\pi$, i.e., every trajectory of the closed-loop system that begins at some $\mathbf{x} \in \mathcal{X}_\pi$ asymptotically approaches the origin. Estimating the ROA is an exhaustive search problem. We arbitrarily sample multiple initial points from the state space $\mathcal{X}$ and then forward simulate the black-box system for a period of time by applying the control law $\pi$. We record the trajectories from different initial points in ROA $\mathcal{X}_\pi$. At any point in time, if the system state goes out of $\mathcal{X}$ for any simulation, then that simulation stops, and the corresponding trajectory is removed from $\mathcal{X}_\pi$. Furthermore, for each trajectory, we record the running average of the Lyapunov energy and if its final value is above some threshold, the corresponding trajectory is removed from $\mathcal{X}_\pi$. 
If the cardinality of $\mathcal{X}_\pi$ is below some threshold, then the control law $\pi$ is classified as invalid. 

\textbf{Bounded actuation.} We limit the magnitude of the control input both during training and evaluation. During training we impose the bounded control in two ways: (i) the input space $\mathcal{U}$ (of the dataset) is bounded (ii) the control input generated by the neural network is also bounded by this limit. In other words, our neural network is already trained to generate bounded control inputs during evaluation. However, we can restrict the control input to an even tighter bound (compared to what was used in training) during evaluation, but that reduces the size of ROA. The actuation bound is determined by the maximum random actuation applied to the system while generating the training dataset such that the system does not leave the training domain.

\textbf{Neural network configuration and training details.} In all experiments, we use $10$K randomly sampled pairs of states and control inputs of the dynamics to train our neural networks, and another $5$K randomly sampled pairs for validation. Networks are trained using Adam optimizer in mini-batches of $32$ samples for $300$ epochs starting with a learning rate of $0.001$, downscaled at every epoch by $0.99$. 

For all simulation examples and all neural networks ($\text{NN}_{\hat g}$, $\text{NN}_{\mathbf{P}}$, and $\text{NN}_{\pi}$), we use multilayer perceptrons (MLPs) with three hidden layers each having $64$ neurons with ReLU activation. The number of neurons in the input and output layers depends on the dimension of the state vector of the specific system. We apply bound on the output of $\text{NN}_{\pi}$ using \textit{tanh} activation.

\textbf{Baselines for comparisons.} We compare the estimated ROA and closed-loop response for the control law obtained from the proposed method with baseline controllers obtained using LQR and RL. 
For the RL baseline, we use the proximal policy optimization algorithm \cite{schulman2017proximal} in actor-critic framework \cite{sutton2018reinforcement}. 
We train the RL agent for 3000 episodes or less if the cost function converges before that. Each episode can have 200 steps at most and can end early if the system goes beyond the allowed domain. To induce randomness, at each step, we store the transition (using the current policy) in a buffer of size 1000 and when the buffer is filled, we randomly sample batches to update the policy. The buffer is then emptied, and we repeat the process. 

\begin{figure*}[t]
  \centering
  \includegraphics[width=0.77\linewidth]{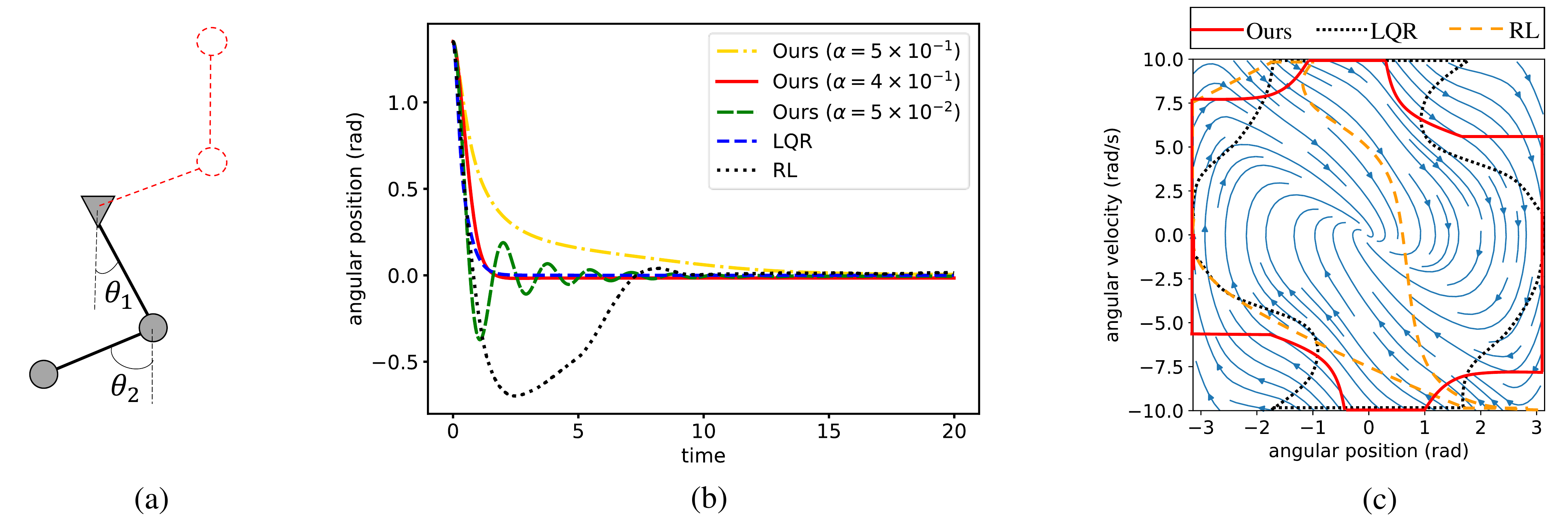}
  \vspace{-10pt}
  \caption{(a) Schematic diagram of the double pendulum model. (b) Example closed-loop responses of angular position of the second link. (c) Phase portrait of the second link subject to the learned control law by our method. ROAs corresponding to the control laws obtained using different methods are shown in different legends.}
  \vspace{-5pt}
  \label{fig:2link}
\end{figure*}

\begin{figure*}[t]
  \centering
  \includegraphics[width=0.8\linewidth]{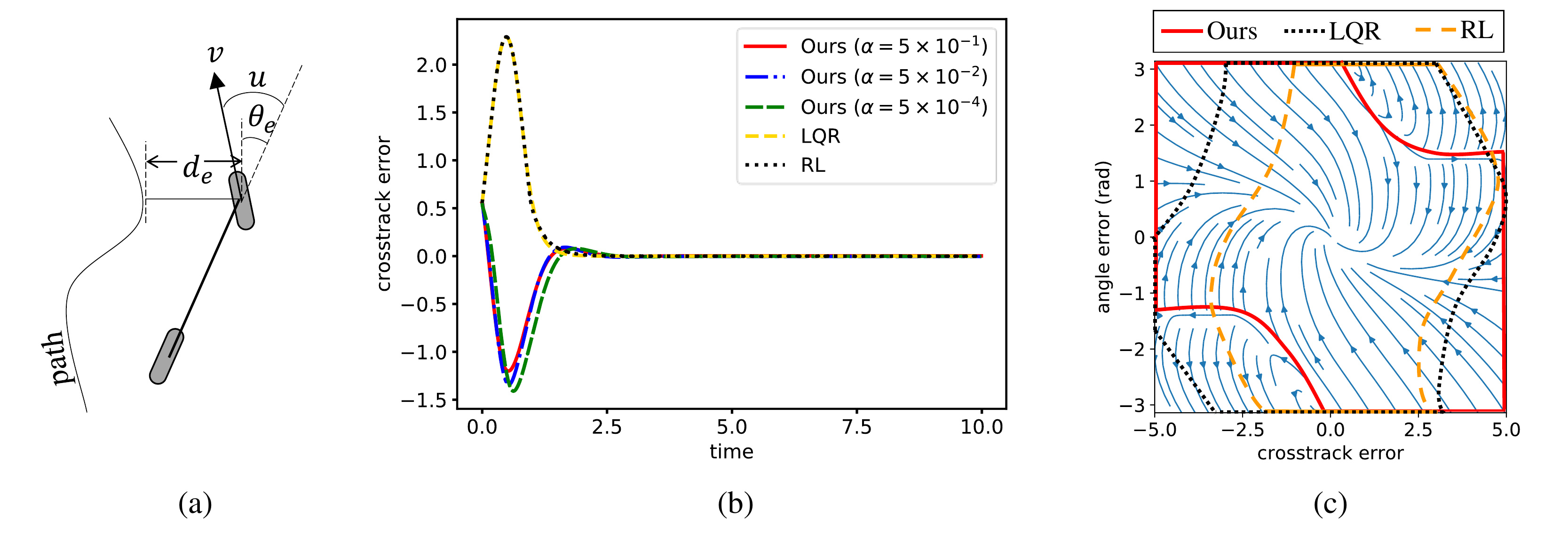}
  \vspace{-10pt}
  \caption{(a) Schematic diagram of the kinematic wheeled vehicle model. (b) Example closed-loop responses of crosstrack error. (c) Phase portrait of the closed-loop system subject to the learned control law by our method. ROAs corresponding to the control laws obtained using different methods are shown in different legends.} 
  \vspace{-5pt}
  \label{fig:vehicle}
\end{figure*}

\begin{figure*}[t]
  \centering
  \includegraphics[width=0.8\linewidth]{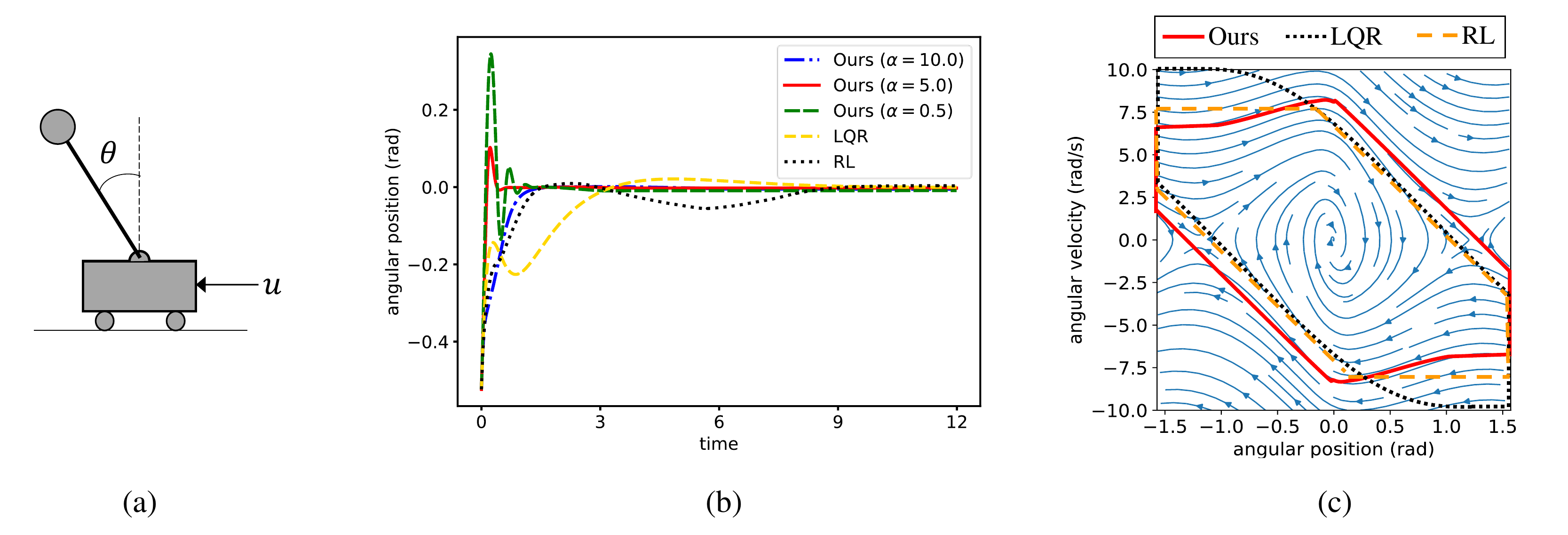}
  \vspace{-10pt}
  \caption{(a) Schematic diagram of an inverted pendulum on a cart. (b) Example closed-loop responses of angular position of the pendulum. (c) Phase portrait of the pendulum subject to the learned control law by our method. ROAs corresponding to the control laws obtained using different methods are shown in different legends.}
  \vspace{-5pt}
  \label{fig:pcart}
\end{figure*}

\begin{figure}[t]
  \centering
  \includegraphics[width=0.75\linewidth]{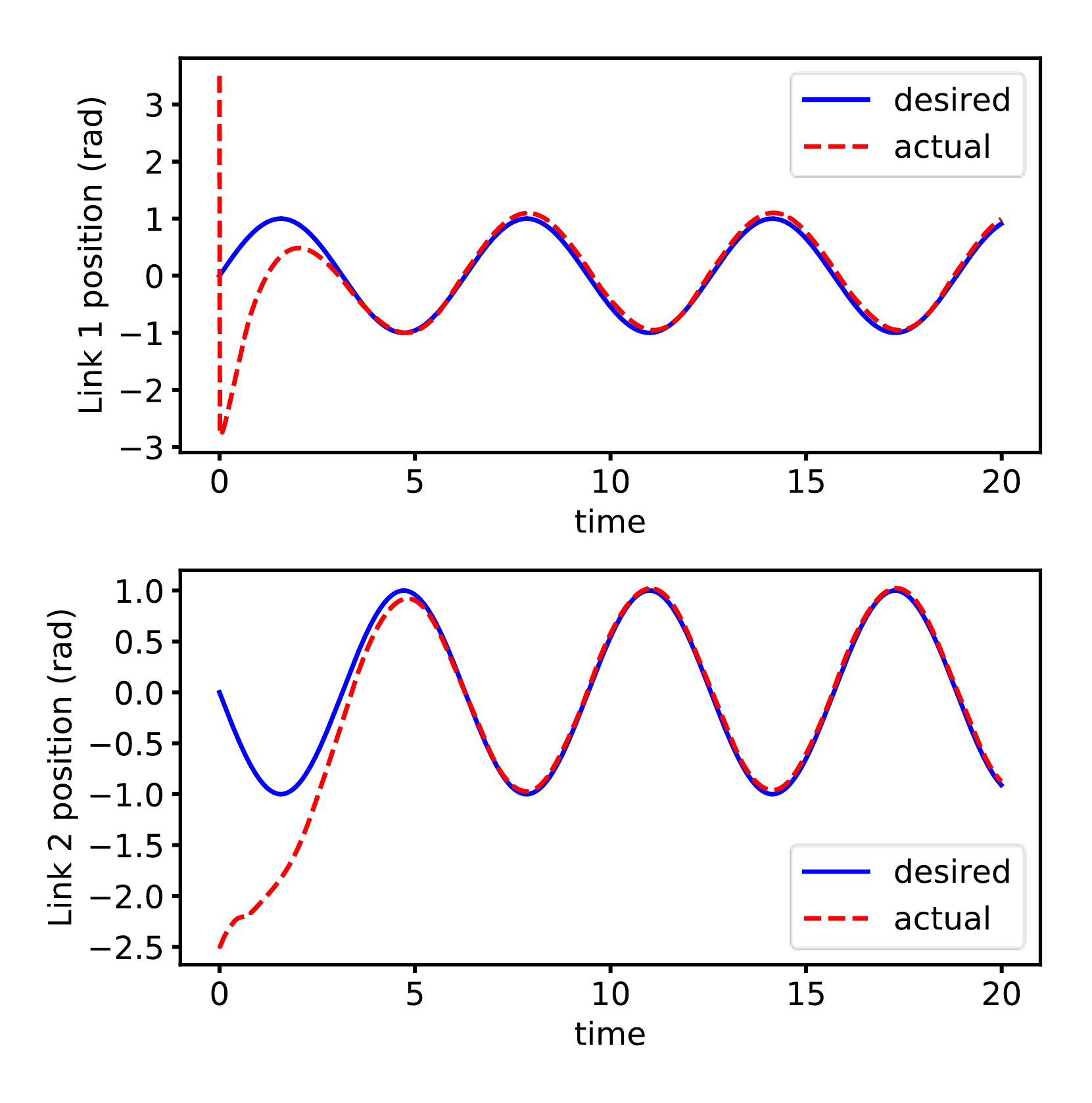}
  \vspace{-5pt}
  \caption{Comparison of trajectories obtained using our controller with their references for the 2-link pendulum tracking problem.}
  \label{fig:manipulator}
  \vspace{-5pt}
\end{figure}

\subsection{Simulation examples}
\textbf{n-Link pendulum balancing and trajectory tracking.}
We consider the problem of balancing an n-link pendulum at a desired posture. The state of the system can be described by the vector $[\theta_1, \cdots \theta_n, \omega_1, \cdots \omega_n]^T$, where $\theta_i$ is the angular position of link $i$ with respect to its target posture and $\omega_i$ is its angular velocity. The system can be moved to the desired posture by applying control input at each pendulum ($n$ control inputs). We enforce a maximum limit on the magnitude of the control input $\|\mathbf{u}\| \leq \bar u = 10 \sqrt{n}$, such that if the system enters the unsafe region, it cannot recover. 
Trajectories for training are generated by the symbolic algebra solver SymPy, using simulation code adapted from \cite{vanderplas_2017}. 
We use $\alpha=0.5$ and the following matrices for $\mathbf{Q}$.
\begin{align}
    \mathbf{Q}_{2\text - link} &= \text{diag}(0.60, 0.32, 0.045, 0.035) \nonumber \\
    \mathbf{Q}_{3\text - link} &= \text{diag}(0.50, 0.35, 0.20, 0.001, 0.001, 0.001)
    \label{eqn:Q_link}
\end{align}

In addition to the balancing problem, we also consider a trajectory tracking problem for the 2-link pendulum where both the links are required to follow two reference trajectories. For this problem, the system state includes an additional error state (with respect to the reference trajectory). 
Same value of $\alpha$ is used, whereas $\mathbf{Q}$ is adjusted to have the same values as (\ref{eqn:Q_link}) in the diagonal elements that correspond to the error state. 

\textbf{Wheeled vehicle path tracking.}
We consider path tracking control of a wheeled vehicle assuming a kinematic vehicle model \cite{park2011smooth}. Path tracking error state can be described by the vector $[d_e, \theta_e]^T$, where $d_e$ is the crosstrack error, measured from the center of the front axle to the nearest path point and $\theta_e$ is the heading error with respect to the tangent at the nearest path point. The desired path needs to be tracked by controlling the steering angle. We enforce a maximum limit on the magnitude of the steering angle $|u| \leq \bar u = \frac{\pi}{6}$. Trajectories for training are generated by the symbolic algebra solver SymPy. 
We use $\alpha=0.05$ and $\mathbf{Q} = \text{diag} (0.96, 0.04)$. 

\textbf{Pendulum on cart balancing.}
Balancing an inverted pendulum upright on a laterally sliding cart is a classic nonlinear control problem. The state of the system can be described by the vector $[x, \theta, v, \omega]^T$, where $x$ and $v$ are the lateral position and velocity, respectively, of the cart, and $\theta$ and $\omega$ are the angular position and velocity, respectively, of the pendulum. The pendulum needs to be stabilized at the upright posture by applying a control input to the cart. We enforce saturation constraints on the cart position and velocity, and a maximum limit on the magnitude of the lateral input force $|u| \leq \bar u = 50$.
Trajectories for training are generated by the symbolic algebra solver SymPy, using simulation code adapted from \cite{moore2013}. 
We use $\alpha=0.5$ and $\mathbf{Q} = \text{diag} (0.0001, 1.0, 0.0001, 0.004)$. 

\subsection{Results and discussion}

\textbf{Analysis of the closed-loop response.} 
Figure \ref{fig:2link}(b), \ref{fig:vehicle}(b), and \ref{fig:pcart}(b) show the closed-loop responses obtained using different methods for n-link pendulum balancing, wheeled vehicle path tracking, and pendulum on cart balancing, respectively. For the pendulum on cart example, We observed that the pendulum attains the desired posture, but the state of cart does not goes to the origin. Properties of the closed-loop response (e.g. overdamped vs underdamped, overshoot, settling time, etc.) for our learned controller can be adapted by tuning the hyperparameter $\alpha$. The impact of $\alpha$ is significant for both pendulum examples, whereas closed-loop responses of wheeled vehicle path following for different values of $\alpha$ are very similar. 

Figure \ref{fig:manipulator} shows the desired trajectories and the trajectories obtained using our controller for 2-link pendulum tracking problem.

\textbf{ROA analysis.}
Figure \ref{fig:2link}(c), \ref{fig:vehicle}(c), and \ref{fig:pcart}(c) compare the ROA obtained using different methods for n-link pendulum balancing, wheeled vehicle path tracking, and pendulum on cart balancing, respectively. Our method attains larger or comparable ROA to the other methods. The phase portraits shown in these figures are obtained using our method and show that the trajectories within the ROA approach the origin. 

\begin{figure}[t]
  \centering
  \includegraphics[width=0.68\linewidth]{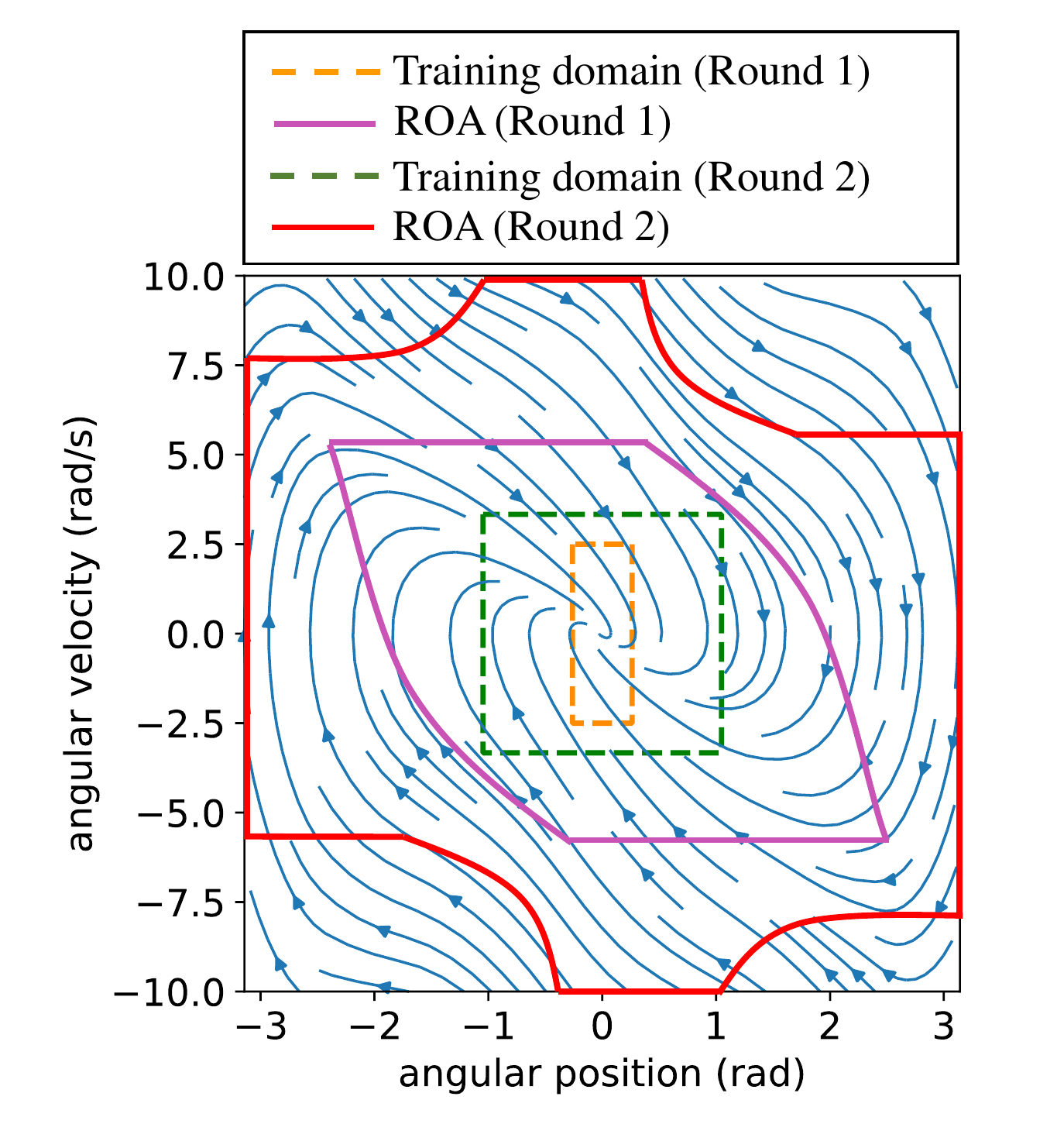}
  \vspace{-5pt}
  \caption{Iterative training for 2-link pendulum balancing. We start training with data collected within a small region around the equilibrium point (Round 1). The learned controller in the first round is used to collect data from a subset of the corresponding ROA. The new data is then used for a second round of training.}
  \label{fig:roa_expansion}
  \vspace{-5pt}
\end{figure}

\textbf{Safe learning from limited data.}
One limitation of the proposed method is that it assumes a significant amount of data from the underlying system is available for training. For safety-critical systems, collecting a large amount of data, particularly from a broad region of operation, is itself a challenging problem. To address this problem, we propose an iterative approach based on the observation that given a small safe region around the equilibrium point, our method learns a controller that yields a ROA beyond the training domain. We start with collecting training data from a known small safe region around the equilibrium point and learn an initial controller. Next, we use the learned controller to collect data over a wider region to update the controller.
\textbf{Algorithm 2} delineates the iterative learning process. We show this iterative process for the 2-link pendulum example in Figure \ref{fig:roa_expansion}.  

\begin{algorithm}
  \caption{Iterative Learning}
  \begin{algorithmic}[1]
     \State  \textbf{input}: Black-box controlled dynamical system $f$, Initial safe state space $\mathcal{X}_0$ and input space $\mathcal{U}_0$ 
     \State \ 
     \State $\mathcal{X}_s \leftarrow \mathcal{X}_0$ \Comment{Initialize safe state space}
     \State $\mathcal{U}_s \leftarrow \mathcal{U}_0$
     \Comment{Initialize safe input space}
     \Repeat 
     \State $\text{NN}_{\pi}$ $\leftarrow$ \textbf{Algorithm 1} ($f$, $\mathcal{X}, \mathcal{U}$) \Comment\parbox[t]{.3\linewidth}{Learn a control law using \textbf{Algorithm 1}} 
     \vspace{2mm}
     \State \text{Compute the ROA $\mathcal{X}_{\pi}$ of controller $\text{NN}_{\pi}$}
     \State $\mathbf{x} \sim p_{\mathbf{x}}(\mathcal{X}_{\pi})$ \Comment Sample batch of states from $X_{\pi}$
     \State $\mathbf{u} \leftarrow \text{NN}_{\pi}(\mathbf{x})$ \Comment Get control inputs using $\text{NN}_{\pi}$
     \State Update $\mathcal{X}, \mathcal{U}$ with new data $\mathbf{x}, \mathbf{u}$
     \Until convergence 
  \end{algorithmic}
\end{algorithm}

\begin{figure}[t]
  \centering
  \includegraphics[width=0.75\linewidth]{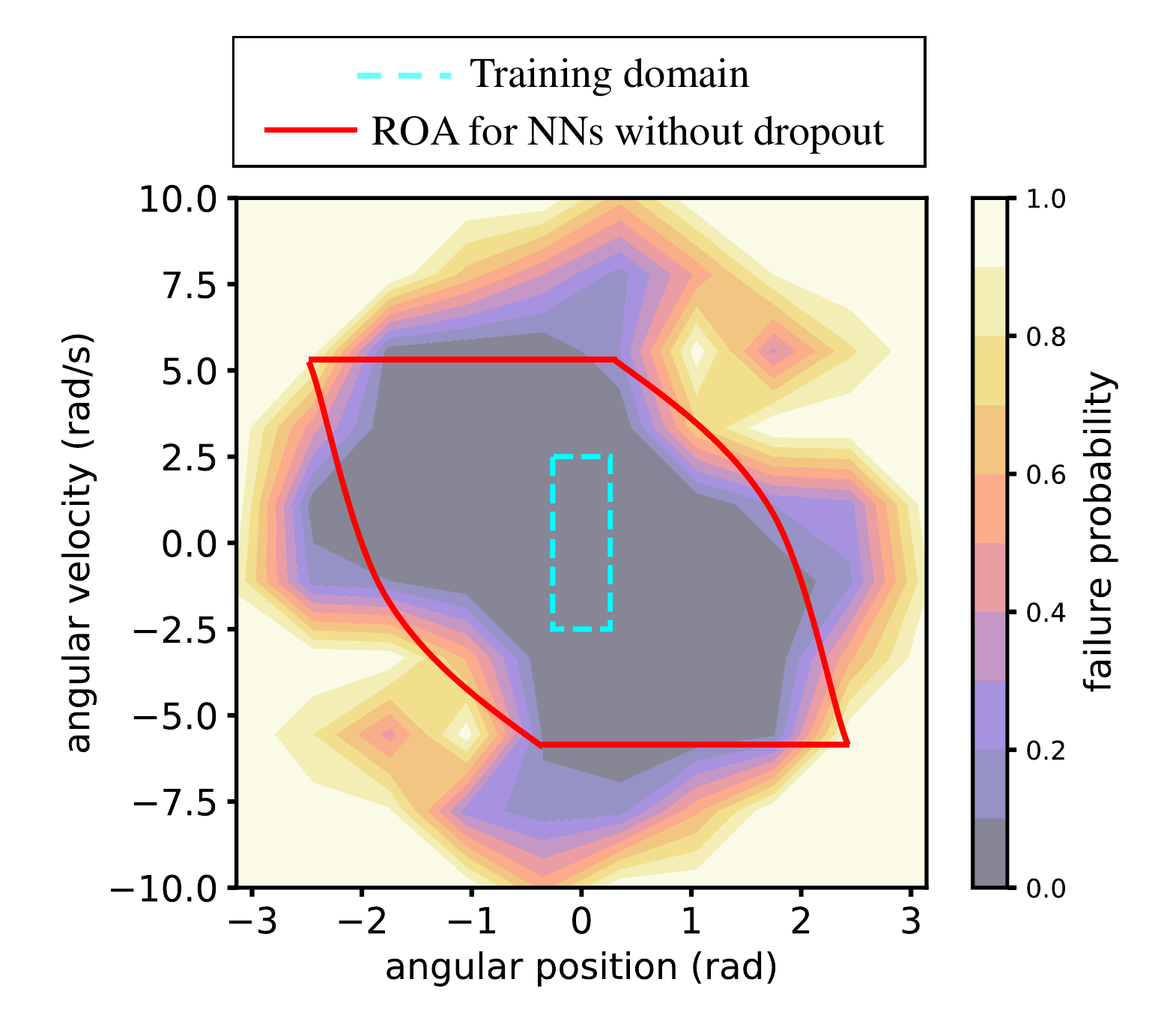}
  \vspace{-5pt}
  \caption{Failure probability in ROA of 2-link pendulum balancing for $50$ MC dropout simulations.}
  \vspace{-5pt}
  \label{fig:uncertainty}
\end{figure}

\textbf{Robustness analysis for model uncertainty.}  
Learning with neural network is subject to inaccuracies due to insufficient data. 
Generally, a Monte Carlo (MC) dropout inference technique is used to capture the uncertainty of neural network output where the neural networks are trained with dropout before every weight layer, and at the test time, multiple inferences are obtained for same input using MC simulations to quantify uncertainty \cite{kendall2017uncertainties}. We use this MC dropout method to analyze the robustness and quantify uncertainties of the learned controller for the 2-link pendulum balancing problem. We train the neural networks with dropout (probability of dropping each hidden neuron is $0.2$) using data collected from a small region around the equilibrium (same as round 1 of \textbf{Algorithm 2}). At test time, for each random initial point, we perform 50 MC simulations of NNs using dropout. We count the number of times trajectory starting from an initial point fails to reach the origin and use the failure probability as the measure of uncertainty at that point. 
Figure \ref{fig:uncertainty} shows that the learned controller is robust under model uncertainty for a significant region outside the training domain.

\textbf{Depth of neural networks.} We use same depth (number of layers) of MLPs for all examples. However, network depth can be increased or decreased depending on the complexity of the system dynamics. For example, wheeled vehicle dynamics is relatively less complex than the other two cases and MLPs having only two layers
shows a very similar ROA, whereas ROA for 2-link pendulum reduces slightly with two-layer MLPs.

\section{Conclusion and Future Work}
\label{sec:conclusion}
We have proposed a novel method for learning control for an unknown nonlinear dynamical system by formulating a system identification task. The proposed method jointly learns a control law to stabilize an unknown nonlinear system and corresponding stable closed-loop dynamics hypothesis. We have demonstrated our approach on various nonlinear control simulation examples. 

The proposed approach assumes a known initial training region from where trajectory data can be collected with zero and random control input. However, such data collection may not be practical in many cases. We plan to address this issue in future work. Developing a method to determine the initial region for an unknown system is also a challenging future work. In this work, we used dropout to ensure the robustness of our controller. However, in future work, we plan to investigate this with thorough robustness analysis, to verify if this approach indeed results in a robust controller. 

\vspace{2cm}
\bibliographystyle{IEEEtran}
\bibliography{ref}


\end{document}